\documentclass[aps,prl,twocolumn,showpacs,superscriptaddress,groupedaddress]{revtex4}  % for review and submission
\usepackage{graphicx}  % needed for figures
\usepackage{dcolumn}   % needed for some tables
\usepackage{bm}        % for math
\usepackage{amssymb}   % for math

\begin{document}

% the following line is for submission, including submission to the arXiv!!
%\hspace{5.2in} \mbox{Fermilab-Pub-04/xxx-E}

\title{The symmetry of the energy momentum tensor does not necessarily reflect the spacetime symmetry: a viscous axially symmetric cosmological solution}

\author{Fatemeh Bagheri}
\affiliation{Department of Physics, Sharif University of Technology, P.O. Box 11155-9161, Tehran, Iran}
\author{Reza Mansouri}
\affiliation{Department of Physics, Sharif University of Technology, P.O. Box 11155-9161, Tehran, Iran}
\affiliation{ School of Astronomy, Institute for Research in Fundamental Sciences (IPM), Tehran, Iran}
       % D0 authors (remove the first 3 lines
                             % of this file prior to submission, they
                             % contain a time stamp for the authorlist)
                             % (includes institutions and visitors)
%\date{\today}

\begin{abstract}
Applying the method of conformal metric to a given static axially symmetric vacuum solution of the Einstein equations and letting the cosmic fluid to be imperfect, we have found the axially symmetric solutions tending to FLRW at space infinity. In fact, The solution represents an axially symmetric spacetime leading to a spherically symmetric Einstein tensor. One should note that this solution is not a vacuum solution such as Kerr metric; this is a cosmological solution which means the matter spreads all over the spacetime. Therefore, we have found a cosmological (non-vacuum) solution of Einstein equations representing a spherically symmetric matter distribution corresponding to a spacetime which does not reflect the same symmetry.

\end{abstract}

\pacs{}
\maketitle
There has recently been an increasing interest to understand cosmological structures within an otherwise expanding universe. Asking for exact solutions representing realistic structures is now far from technical abilities to formulate the problem. The simplest case of spherically symmetric cosmological structures and black holes have already been studied \citep{razbin2014relativistic, firouzjaee2010asymptotically, firouzjaee2012we, firouzjaee2012spherical}. As a step further towards a more realistic model we are asking for axially symmetric cosmological structures represented by axially symmetric inhomogeneous solutions of Einstein equations being asymptomatically FLRW. There is an extensive literature on axially symmetric solutions of Einstein equations, some of them on cosmological solutions being axially symmetric \citep{trendafilova2011static, senovilla2007singularity, senovilla1987stationary, davidson1996petrov, davidson1994infinite, evans1977static, stephani2003exact, fernandez1997singularity, senovilla2000cylindrically, davidson1993cylindrically}. Cosmological solutions with a perfect fluid being axially symmetric have been discussed in \citep{dandach1984axially}. There, a family of axially symmetric metrics generalizing the spherically symmetric analogs has been found depending on one extra arbitrary function of $r$ and $\theta$ (as compared to the spherically symmetric case).\\
In general, axial symmetry is defined in terms of an isometric $SO_2$ mapping of space-time which maps a set of fixed points to a time-like two dimensional surface. The plane is called the surface of symmetry and the set of the fixed points is called the axis of rotation. This is accomplished by the existence of a space-like Killing vector field $\eta =  \frac{\partial}{\partial\phi}$ with closed (compact) trajectories vanishing on the axis of rotation. The existence of a second space-like Killing vector makes the space-time cylindrically symmetric.\\
%Assuming now an axially symmetric space-time, one may introduce coordinates such that the metric coefficients are independent of the coordinate '[16]. Therefore, the axially symmetric space-time may be written in spherical coordinates in the following form
%\begin{equation}
 % ds^2 = - F(r, \theta, T)^2 dT^2 + A(r, \theta, T)^2 dr^2 + B(r, \theta, T)^2 d\theta^2 + D(r, \theta, T)^2 d\phi^2.
%\end{equation}
In the stationary case we may then assume the existence of a timelike Killing vector $\xi = \frac{\partial}{\partial t}$ and introduce a coordinate system such that two of the spacetime dimensions are tangent to the Killing vectors $\xi = \frac{\partial}{\partial t}$ and $\eta = \frac{\partial}{\partial \phi}$. The stationary axially symmetric metric may then be written as
\begin{eqnarray}
  ds^2 = e^{-2U}[\gamma_{MN}dx_M dx_N +W^2 d\phi^2]- e^{2U}(dt + Ad\phi)^2,
\end{eqnarray}
where $U$,$\gamma_{MN}$, $W$, and $A$ are functions of coordinates $x^M = (x^1, x^2)$. Without loss of generality we may use an isotropic coordinate system, the so-called Weyl’s canonical coordinates, such that
\begin{eqnarray}
  \gamma_{MN} = e^{2k}\delta_{MN}.
\end{eqnarray}
The resulting form of the metric will then be
\begin{eqnarray}
  ds^2 = e^{-2U}[e^{2k}(d\rho^2 + dz^2) + \rho^2 d\phi^2] - e^{2U}(dt + Ad\phi)^2.
\end{eqnarray}
The metric is static if the two Killing vectors are perpendicular, i.e. $A$ vanishes:
\begin{eqnarray}
  ds^2 = e^{-2U}[e^{2k}(d\rho^2 + dz^2) + \rho^2 d\phi^2] - e^{2U}dt^2,
\end{eqnarray}
with $U$ and $k$ depending only on $\rho$ and $z$ \citep{stephani2003exact}. In the static vacuum case there are solutions in closed form in spherical coordinates with
\begin{eqnarray}
  U = \Sigma_{n = 0}^{\infty}a_n r^{-(n+1)}P_n(\cos\theta),
\end{eqnarray}
and
\begin{eqnarray}
  k = -\Sigma_{n = 0}^{\infty}\frac{a_l a_m(l + 1)(m + 1)}{(l + m + 2)r^{l+m+2}}(P_lP_m - P_{l+1}P_{m+1}),
\end{eqnarray}
where $P_n(\cos\theta)$ are Legendre functions. For the simplest case with $n = 0$, we obtain
\begin{eqnarray}
  U = - \frac{m}{r}, \quad k = -\frac{m^2\sin\theta^2}{r^2}.
\end{eqnarray}
Stationary axially symmetric solutions with perfect fluid are also discussed in \citep{senovilla1987stationary, davidson1996petrov, davidson1994infinite}. For this more general case with the metric in the form (4), the fluid velocity may be written as a linear combination of the two Killing vectors:
\begin{eqnarray}
  u^{[a}\xi^{b}_{A}\xi^{c]}_{B} = 0.
\end{eqnarray}
Therefore, there exists a 2-surface orthogonal to the group orbits. The velocity is then necessarily non-expanding, i.e. $u_{a;}^{a} = 0$. Should the angular velocity be constant, we obtain
\begin{eqnarray}
  \sigma = \theta = 0 \Leftrightarrow u_{(a;b)} + u_{(a}\dot{u}_{b)} = 0,
\end{eqnarray}
leading to a rigidly rotating fluid. However, asking for matter-dominated FLRW at space infinity as a boundary condition, one can show that these solutions reduce in general to the inhomogeneous spherically symmetric LTB solutions with the simple FLRW solutions as a special case. It is also simple to see that non of the other solutions discussed in \citep{fernandez1997singularity}, \cite{senovilla2000cylindrically}, and \cite{davidson1993cylindrically} approaches to FLRW as r approaches to infinity. The stationary solutions with axial symmetry (or cylindrical symmetry) in \cite{stephani2003exact} are non-expanding, and hence may not be used as cosmological solutions. We, therefore, look for different methods to find axially symmetric cosmological solutions with perfect fluid approaching FLRW at infinity.\\
Two methods of generating new metrics from existing ones have been introduced in literature: using minimally coupled massless scalar field and the use of conformal transformations. Here we intend to start again with a vacuum static solution and generate an expanding cosmological solution by using a conformal transformation. Consider the Einstein tensor with non-zero off-diagonal entries, corresponding to an imperfect fluid with an energy-momentum tensor with viscous terms \citep{carot1986conformal} and \citep{tupper1990conformally}. We use the simplest asymptotically flat case of Weyl’s metric (5) as a vacuum solution in the spherical coordinates:
\begin{eqnarray}
  ds^2 =&&e^{\frac{2m}{r}}e^{\frac{-m^2 \sin^2\theta}{r^2}}(dr^2 + r^2d\phi^2)\\ \nonumber
  &+&e^{\frac{2m}{r}} r^2 \sin^2 \theta d\phi^2\\ \nonumber
  &-&e^{\frac{-2m}{r}}dt^2.
\end{eqnarray}
The solution has the Petrov type I and the Segre type [(1,111)]. Applying now a conformal transformation with $\Omega(t) = e^{2\omega(t)}$, we obtain
\begin{eqnarray}
  e^{2\omega(t)}ds^2 =&&e^{2\omega(t)+\frac{2m}{r}}e^{\frac{-m^2 \sin^2\theta}{r^2}}(dr^2 + r^2d\phi^2)\\ \nonumber
  &+& e^{2\omega(t)+\frac{2m}{r}}r^2 \sin^2 \theta d\phi^2\\ \nonumber
  &-& e^{2\omega(t)-\frac{2m}{r}}dt^2.
\end{eqnarray}
Assuming $\omega_{,t} = \dot{\omega}(t) > 0$, the fluid velocity is given by
\begin{eqnarray}
  u_\mu = -\frac{1}{\lambda}\dot{\omega}(t)\delta_{\mu}^{0},
\end{eqnarray}
where $\lambda = \dot{\omega}(t)e^{-\omega(t)+ \frac{m}{r}}$. The expansion scalar, density, pressure, and the heat conduction are calculated to be
\begin{eqnarray}
  \Theta = 3\dot{\omega}(t)e^{-\omega(t)+\frac{m}{r}},
\end{eqnarray}
\begin{eqnarray}
  \rho = 3\dot{\omega}^2(t)e^{-2\omega(t)+\frac{2m}{r}},
\end{eqnarray}
\begin{eqnarray}
  p = -[2\ddot{\omega}(t) + \dot{\omega}^2(t)]e^{-2\omega(t)+\frac{2m}{r}},
\end{eqnarray}
\begin{eqnarray}
  q_{\mu} = -2\frac{2m}{r^2}\dot{\omega}(t)e^{-\omega(t)+\frac{m}{r}}\delta_{\mu}^{1}.
\end{eqnarray}
The shear- and pressure-tensor are given by $\pi_{\mu\nu} = 2\lambda \sigma_{\mu\nu} = 0$. Note that the density is positive and the shear is zero. In this case the Segre type for the energy tensor is changed to [2 11]. As we are interested in a solution being asymptotically FLRW, we take the pressure to be zero. This is achieved by assuming the conformal factor to be
\begin{eqnarray}
  \omega(t) = 2\ln(\frac{at}{2} + \frac{b}{2})~,
\end{eqnarray}
where $a$ and $b$ are two arbitrary constants. Considering $a = 2$ and $b = 0$ the change of variable $T = t^3/3$ leads to
\begin{eqnarray}
  ds^2 = &&a^2(T)e^{\frac{2m}{r}} e^{-\frac{m^2 \sin^2\theta}{r^2}}(dr^2 + r^2 d\theta^2)\\ \nonumber
  &+& a^2(T)e^{\frac{2m}{r}}r^2 \sin^2\theta d\phi^2\\ \nonumber
  &-& e^{-\frac{2m}{r}} dT^2,
\end{eqnarray}
where $a(T ) = (3T )^{2/3}$. The density and the heat conduction are then given by
\begin{eqnarray}
  \rho = \frac{4}{3T^2}e^{\frac{2m}{r}},
\end{eqnarray}
and
\begin{eqnarray}
  q_{\mu} = -\frac{4}{3mr^2(3T)} e^{\frac{2m}{r}}\delta_{\mu}^{1}.
\end{eqnarray}
At large $r$ the heat conduction approaches to zero, and the metric approaches a dust dominated FLRW universe with the expected density behavior. We have therefore arrived at an axially symmetric cosmological solution of Einstein equations with an imperfect fluid which is asymptotically FLRW. We are tempted to assume that this solution represents an axially symmetric overdense structure within an FLRW universe.\\
Here we have found an axially symmetric solution of Einstein equations corresponding to a spherically symmetric matter distribution. Although the metric is axially symmetric the corresponding Ricci- and Einstein-tensors are spherically symmetric. A fact opposing our expectation since it has always been assumed in general relativity that the symmetry of the matter tensor is reflected in the symmetry of spacetime expressed in the metric tensor (see for example the discussion in \cite{hall2004symmetries}). To be sure that this metric is not spherically symmetric, we calculate the Ricci and the Kretschmann scalars:
\begin{eqnarray}
  R = 6e^{2\frac{m}{r} -2\omega(t)}(\ddot{\omega} + \dot{\omega}^2),
\end{eqnarray}
\begin{eqnarray}
  R_{\mu\nu}R^{\mu\nu} =&&e^{\frac{4m}{r}-4\omega(t)}[9\ddot{\omega}^2 + 3(\ddot{\omega} + 2\dot{\omega}^2)^2]\\ \nonumber
  &-& 2[\frac{2m}{r^2}\dot{\omega}(t)]^2 e^{-4\omega(t) + \frac{m^2 \sin^2\theta}{r^2}},
\end{eqnarray}
\begin{eqnarray}
  K = &&\frac{e^{-4\omega(t)}}{r^{10}} e^{\frac{2m^2 \sin^2\theta}{r^2}-\frac{4m}{r}}\big(16 m^6 \sin^2\theta \\ \nonumber
  &+& 48 m^5 r \sin^2\theta + 48 m^4 r^2 (\sin^2\theta + 1) \\ \nonumber
  &-& 96 m^3 r^3 + 48 m^2 r^4\big) \\ \nonumber
  &-&e^{-4\omega(t)}\big(\frac{16m^2}{r^4}e^{\frac{m^2 \sin^2\theta}{r^2}}\dot{\omega}^2 + 12 e^{}(\dot{\omega}^4 + \ddot{\omega}^2)\big).
\end{eqnarray}
Although the Ricci scalar is independent of the angle $\theta$, obviously the Kretschmann scalar and the square of the Ricci tensor are angular dependents. Therefore, the spacetime is not spherically symmetric in contrast to the spherically symmetric matter distribution. This can also be seen by the fact that the Lie algebra of the corresponding spherical Killing vectors is not closed. This is to our knowledge the first case of a space-time having a symmetry different from that of the matter source. \\
This may also be stated explicitly in the form that in the Einstein gravity the symmetry of the matter distribution does not necessarily reflect the spacetime symmetry. To force the space-time to have the same symmetry as the matter distribution may lead to an action different from that of Einstein-Hilbert. We may as an example take a term proportional to the square of the Ricci tensor or the Kretschman scalar and take the Lagrangian for the gravitational field to be proportional to $\alpha R + \beta R_{\mu\nu}R^{\mu\nu} + \gamma K$. A novel feature worth to consider!\\
The solution we have found has a very novel feature: although the spacetime is axially symmetric, the corresponding Einstein tensor and consequently the energy momentum of the cosmic fluid is spherically symmetric. The important point here is that this solution is not a vacuum solution such as Kerr metric; \textbf{this is a cosmological solution}, i.e. the cosmological fluid is nonzero for every point in the spacetime. In other words, matter is not a singularity such as in Kerr solution. To our knowledge, this is the first time a non-vacuum solution of the Einstein equations has been reported such that the spacetime symmetry does not reflect the matter symmetry. We have, therefore, shown on hand of a very specific example that the symmetry of matter distribution in the Einstein equations does not necessarily reflect the spacetime symmetry.

\bibliography{main}
%\begin{thebibliography}{99}
%\end{thebibliography}

\end{document}